\documentclass[12 pt, amsfonts, amssymb,color]{article}

\usepackage{amsmath,amssymb,amsfonts,latexsym}

\begin{document}

\renewcommand\theequation{\arabic{section}.\arabic{equation}}
\catcode`@=11 \@addtoreset{equation}{section}
\newtheorem{axiom}{Definition}[section]
\newtheorem{axiom1}{Theorem}[section]
\newtheorem{axiom2}{Example}[section]
\newtheorem{axiom3}{Lemma}[section]
\newtheorem{prop}{Proposition}[section]
\newtheorem{axiom5}{Corollary}[section]
\newcommand{\be}{\begin{equation}}
\newcommand{\ee}{\end{equation}}

\newcommand{\equal}{\!\!\!&=&\!\!\!}
\newcommand{\rd}{\partial}
\newcommand{\g}{\hat {\cal G}}
\newcommand{\bo}{\bigodot}
\newcommand{\res}{\mathop{\mbox{\rm res}}}
\newcommand{\diag}{\mathop{\mbox{\rm diag}}}
\newcommand{\Tr}{\mathop{\mbox{\rm Tr}}}
\newcommand{\const}{\mbox{\rm const.}\;}
\newcommand{\cA}{{\cal A}}
\newcommand{\bA}{{\bf A}}
\newcommand{\Abar}{{\bar{A}}}
\newcommand{\cAbar}{{\bar{\cA}}}
\newcommand{\bAbar}{{\bar{\bA}}}
\newcommand{\cB}{{\cal B}}
\newcommand{\bB}{{\bf B}}
\newcommand{\Bbar}{{\bar{B}}}
\newcommand{\cBbar}{{\bar{\cB}}}
\newcommand{\bBbar}{{\bar{\bB}}}
\newcommand{\bC}{{\bf C}}
\newcommand{\cbar}{{\bar{c}}}
\newcommand{\Cbar}{{\bar{C}}}
\newcommand{\Hbar}{{\bar{H}}}
\newcommand{\cL}{{\cal L}}
\newcommand{\bL}{{\bf L}}
\newcommand{\Lbar}{{\bar{L}}}
\newcommand{\cLbar}{{\bar{\cL}}}
\newcommand{\bLbar}{{\bar{\bL}}}
\newcommand{\cM}{{\cal M}}
\newcommand{\bM}{{\bf M}}
\newcommand{\Mbar}{{\bar{M}}}
\newcommand{\cMbar}{{\bar{\cM}}}
\newcommand{\bMbar}{{\bar{\bM}}}
\newcommand{\cP}{{\cal P}}
\newcommand{\cQ}{{\cal Q}}
\newcommand{\bU}{{\bf U}}
\newcommand{\bR}{{\bf R}}
\newcommand{\cW}{{\cal W}}
\newcommand{\bW}{{\bf W}}
\newcommand{\bZ}{{\bf Z}}
\newcommand{\Wbar}{{\bar{W}}}
\newcommand{\Xbar}{{\bar{X}}}
\newcommand{\cWbar}{{\bar{\cW}}}
\newcommand{\bWbar}{{\bar{\bW}}}
\newcommand{\abar}{{\bar{a}}}
\newcommand{\nbar}{{\bar{n}}}
\newcommand{\pbar}{{\bar{p}}}
\newcommand{\tbar}{{\bar{t}}}
\newcommand{\ubar}{{\bar{u}}}
\newcommand{\utilde}{\tilde{u}}
\newcommand{\vbar}{{\bar{v}}}
\newcommand{\wbar}{{\bar{w}}}
\newcommand{\phibar}{{\bar{\phi}}}
\newcommand{\Psibar}{{\bar{\Psi}}}
\newcommand{\bLambda}{{\bf \Lambda}}
\newcommand{\bDelta}{{\bf \Delta}}
\newcommand{\p}{\partial}
\newcommand{\om}{{\Omega \cal G}}
\newcommand{\ID}{{\mathbb{D}}}

\title{ Generalized damped Milne-Pinney equation and Chiellini method}
\author{Supriya Mukherjee\footnote {Email: supriyaskbu2013@gmail.com}\\
Department of Mathematics \\ Gurudas college, Kolkata-54. India\\
\and
A.Ghose Choudhury\footnote{E-mail aghosechoudhury@gmail.com }\\
Department of Physics, Surendranath  College,\\ 24/2 Mahatma
Gandhi Road, Calcutta-700009, India.\\
\and
Partha Guha\footnote{E-mail: partha@bose.res.in}\\
S.N. Bose National Centre for Basic Sciences \\
JD Block, Sector III, Salt Lake \\ Kolkata - 700098,  India \\
}

\date{ }

 \maketitle

\begin{abstract}
We adopt the Chiellini integrability method to
find the solutions of various generalizations of the damped Milne-Pinney equations.
In particular, we find the solution of the damped Ermakov-Painlev\'e II equation
and generalized dissipative Milne-Pinney equation.
\end{abstract}

\smallskip

\paragraph{Mathematics Classification (2000)}: 34C14, 34C20.

\smallskip


\section{Introduction}
Ermakov \cite{Ermakov} used the Milne Pinney equation \cite{Milne}, \cite{Pinney} while investigating a first integral for the corresponding time dependent harmonic oscillator. Since then, this nonlinear equation has gained intensive attention \cite{Haas}, \cite{CdL} of physicists and engineers due to its widespread application in many physical problems such as propagation of laser beams in nonlinear media, plasma dynamics etc.
It is well known that the general solution for the Milne-Ermakov-Pinney equation
\be\label{EP}
\ddot{y} + \omega^2(t)y = \frac{\kappa}{y^3},
\ee
where $\ddot{y}$ denotes double differentiation of $y$ with respect to time $t$, $\omega = \omega(t)$ is a time-dependent frequency function and
$\kappa$ is a numerical constant can be written as
$y = (Ax_{1}^{2} + 2Bx_1x_2 + Cx_{2}^{2})^{1/2}$,
where $A, B$ and $C$ are constants such that $AC-B^2 = \kappa$ and $x_1$ and $x_2$ are
two independent solutions for the time-dependent harmonic oscillator equation
$\ddot{x} + \omega^2(t)x = 0$.

Equation (\ref{EP}) does not include any mechanism of damping.  Hence it is natural to add a term linear in the
velocity, yielding the damped Milne-Ermakov-Pinney equation
\be\label{dEP}
\ddot{y} + \mu\dot{y} +  \omega^2(t)y = \frac{\kappa}{y^3},
\ee
where $\mu >0$ is a constant positive parameter. Equation (\ref{dEP}) can be transformed into generalized Emden-Fowler equation of index $-3$ which satisfies integrability.

In recent times a hybrid Ermakov-Painlev\'{e} II system was derived by  Rogers \cite{Rogers} in a
pioneering work as a
reduction of a coupled $N+1$-dimensional
Manakov-type NLS system. He showed that the Ermakov invariants admitted by the hybrid system were key
to its systematic reduction in terms of a single component Ermakov-Painlev\'e II equation which, in
turn, may be linked to the integrable Painlev\'e II equation.

\smallskip

The application of the Chiellini integrability condition to find the solutions of nonlinear differential equations has been recently
promoted by two groups; Harko, Mak and their coauthors \cite{MH1,MH2,MH3,MH4} and Mancas and Rosu \cite{RMC,MR,MR1}. It must be worth to
note that the Chiellini integrability condition appears quite naturally for Hamiltonization of the Li\'enard equation using
the Jacobi multiplier technique.
Our intention in this letter is to extend the scope of the integrability condition given by Chiellini to find out solutions to
new kind of damped Ermakov-Milne-Pinney systems. In particular, we obtain the analytic solutions of
the damped Ermakov-Painlev\'e II equation and generalized damped Milne-Pinney equation.

\smallskip

The main result of this paper is given as follows.
\begin{prop}

(a)\,  Let the damped Ermakov-Painlev\'e II equation $$ \ddot{y} + g(y)\dot{y} + h(y) = 0, \qquad h(y) = \lambda y + \epsilon y^3 - \frac{\eta}{y^3}, $$ satisfies the Chiellini integrability condition $\frac{d}{dy}(\frac{h(y)}{g(y)}) = pg(y),$ for some constant $p$.
The solution of the
above equation is given by
$$
y = \sqrt{\frac{1}{(t-t_{0})^2}+\sqrt{\frac{c}{3}}}, \,\,\,\,  \epsilon = -1 \qquad
y = \sqrt{\bigg(\frac{c^2}{16 \lambda^2}-\frac{\eta}{\lambda}\bigg)^{\frac{1}{2}} \sin[2\sqrt{2\lambda}(t-t_0)]+\frac{c}{4\lambda}}, \,\,\,\, \epsilon =0.
$$
(b)\, If the generalized damped Milne-Pinney equation
$$
\ddot{y}+g(y)\dot{y}+\lambda y =\frac{k_1}{y^3}+\frac{k_2}{y^2}+\sum_{n=0}^R\delta_n y^{2n+1}
$$ satisfies Chiellini condition then a  parametric solution of this equation for $R=0$ is given by
$$
t=y_0\omega+\frac{f^{'}(y_0)}{4\wp^{'}(\omega_0)}[\log\frac{\sigma(\omega+\hat{c}-\omega_0)}{\sigma(\omega+\hat{c}+\omega_0)}
+2(\omega+\hat{c})\zeta(\omega_0)]+\delta
$$
$$
\displaystyle y=y_0+\frac{f^{'}(y_0)}{4[\wp(\omega+\hat{c})-\frac{f^{''}(y_0)}{24}]}
$$
where $\delta$ is an integrating constant and $\hat{c}$ being any fixed constant.
\\Also $f(y)=2(\delta_0-\lambda)y^4+cy^2-4k_2y-2k_1 $, $y_0$ is a root of the equation $f(y)=0$
 and $\wp(\omega)$ is the {\it Weierstrass} $\wp$ - {\it function}
\end{prop}

Rest of the article is devoted to the proof of our main result.

\section{Chiellini method and solution of equations}

The first order Abel differential equation $\cite{AD}$ of the first kind plays an important role in many physical and
mathematical problems. The connection between the second-order nonlinear differential equations and
the Abel equation is well known $\cite{HT}$,$\cite{CM}$ and the solutions to such differential equations can often be obtained
via the solutions of the corresponding Abel differential equations. A second order differential equation
of the  Li$\acute{e}$nard type $\cite{AL}$ given by
\begin{equation}
 \ddot{Y}+ g(Y) \dot{Y}+ h(Y)=0
 \end{equation} may be tranformed into a first-order Abel differential equation of second kind, namely
\begin{equation}z\frac{dz}{dY}+g(Y)z+h(Y)=0\end{equation} by the transformation $ \dot{Y}= z(Y(t)) $, which in turn is transformed
to the Abel equation of first kind
 \begin{equation}
 \frac{dX}{dY}=g(Y) X^2 +h(Y)X^3
  \end{equation}
  via the transformation $z=\frac{1}{X}$. However, the criterion of integrability of such equations greatly depends on the expressions of $g(Y)$ and $h(Y)$. An important observation by Chiellini $\cite{AC}$ in 1931 states that a first kind Abel differential equation (2.3) is exactly integrable if the functions $g(Y)$ and $h(Y)$ satisfies the condition
  \begin{equation}
 \frac{d}{dY} \left(\frac{h(Y)}{g(Y)}\right)=pg(Y)
 \end{equation} for some constant {\it p}.
 This integrability condition has been applied in 1960s by Bandi$\acute{c}$ who wrote a couple of mathematical papers $\cite{IB1}$, $\cite{IB2}$ and then by Borghero and Melis $\cite{FB}$ in the Szebehely's problem. Recently, this integrability condition as gained much attention by Mak and Harko $\cite{MH1},\cite{MH2},\cite{MH3}$ in obtaining general solutions of the first-kind Abel equations from a particular solution. This result has also been used by Yurov and Yurov $\cite{YY}$ in cosmology and again by Harko et al $\cite{MH4}$ in case of particular Li$\acute{e}$nard equations.

\smallskip

The Chiellini condition not only ensures the integrability of a system but it also helps to find the solution.
If we further require that $z= c_k\frac{h(Y)}{g(Y)}$ then its substitution in equation (2.2) leads to
$$ pc_k^2 + c_k+1=0 \Rightarrow c_k = \frac{-1\pm\sqrt{1-4p}}{2p}.$$
 For simplicity we choose $c_k=1$ which gives the value of $p=-2$ in equation (2.4). Thus from $ \dot{Y}= z(Y(t)) $ we have
 \begin{equation}
  \dot{Y}= \frac{h(Y)}{g(Y)}.
  \end{equation}
  Using this result we arrive at a much relevant observation that equation (2.1) can be turned to the non dissipative equation
\begin{equation}
 \ddot{Y}+H(Y)=0, \hskip50pt H(Y)=2h(Y)
 \end{equation}
  where the function $h(Y)$ is scaled up by a factor $2$.
  \\ This result allows us to find the dissipation function in (2.1)without actually knowing $Y$.
Multiplying $\ddot{Y}+2h(Y)=0$ by $\dot{Y}$ and integrating we have
\begin{equation}
\dot{Y}\ddot{Y}+2\dot{Y}h(Y)=0\Rightarrow \dot{Y}^2=-4\int{h(Y)dY} +c
\end{equation} where $c$ is an integrating constant.
\\ Thus from (2.5) we have
\begin{equation}
g(Y)=\frac{h(Y)}{\sqrt{c-4\int{h(Y)dY}}}
\end{equation}
Upon further integration of (2.7) we have
\begin{equation}
t-t_0 = \int{\frac{dY}{\sqrt{c-4\int{h(Y)dY}}}}
\end{equation}
where $t_0$ depends on an initial condition.

\section{Solutions of Dissipative Ermakov-Painlev\'e II and generalized Milne-Pinney equations}

Combining the terms of both Ermakov-Pinney equation and the Painlev\'e II  we obtain the following equation
\be
\ddot{y} + \frac{\tau}{2}y + \epsilon y^3 = -\frac{1}{4y^3}(\gamma - \frac{\epsilon}{2})^2.
\ee
This nonlinear equation is known as the (single component) Ermakov-Painlev\'e II equation and was derived by
Rogers \textit{et al} \cite{Conte,Rogers}. It is related  \cite{Gromak} to the
Painlev\'e II equation
\be
\ddot{z} = 2z^3 + \tau z + \gamma,
\ee
where $$ z = \frac{\epsilon}{2y^2}\big( \gamma - \frac{\epsilon}{2} - 2y\dot{y} \big). $$
If we express $y = \sqrt{|\phi|^2 + \psi|^2}$ then the canonical single component Ermakov-Painlev\'e II equation
yields a particular Ermakov-Ray-Reid system ( for details, see \cite{Rogers}) and admits
the characteristic invariant which may be exploited systematically to construct the solutions.

Let us assume the equation (3.1) as
\begin{equation}
 \ddot{y}+\lambda y+\epsilon {y^3}=\frac{\eta}{y^3}
 \end{equation}
 where $ \lambda = \frac{\tau}{2}$ and $\eta = - \frac{1}{4}(\gamma - \frac{\epsilon}{2})^2 $
 \\If
\begin{equation}
 h(y)=\lambda y+\epsilon {y^3}-\frac{\eta}{y^3}
\end{equation}
\\then equation (3.3) can be written as
\begin{equation}
 \ddot{y} +h(y) =0
\end{equation}
 We introduce the dissipative Ermakov-Painlev\'e II equation having same $h(y)$ as in the non dissipative
case but with an additional damping term. The equation
\begin{equation}
 \ddot{Y}+ g(Y) \dot{Y}+ h(Y)=0
\end{equation}
is called Chiellini dissipative Ermakov Painlev\'e II equation because the damping coefficient $g(Y)$ will be obtained from the Chiellini integrability condition.
Using $h(Y)$ from (3.4) in (2.7)  we have
\begin{equation}
\dot{Y}=\sqrt{c-\epsilon Y^4 -2\lambda Y^2 -2\eta Y^{-2} }
\end{equation}
and in (2.8) we have
\begin{equation}
g(Y)=\frac{\lambda Y^2+\epsilon Y^4 -\eta Y^{-2}}{\sqrt{-\epsilon Y^6 -2\lambda Y^4 +cY^2 -2\eta  }}
\end{equation} Further from (3.7)upon integration once more we have
\begin{equation}
\begin{array}{cc}
 \displaystyle Y^2=\frac{1}{(t-t_{0})^2}+\sqrt{\frac{c}{3}},\hskip120pt & \epsilon = -1 \\
  \displaystyle=\sqrt{\frac{c^2}{16 \lambda^2}-\frac{\eta}{\lambda}}^{\frac{1}{2}} \sin[2\sqrt{2\lambda}(t-t_0)]+\frac{c}{4\lambda}, & \epsilon =0
\end{array}
\end{equation}
$ t_0 $ depending on initial conditions.

\subsection{Generalized dissipative Milne-Pinney equation}

At first we embark a simple equation of this category and obtain its solution.
Let us consider the equation
\begin{equation}
\displaystyle\ddot{Y}+g(Y)\dot{Y}-\frac{\delta}{Y^5}=0
\end{equation}
This may be written as equation(2.1) with $\displaystyle h(Y)=-\frac{\delta}{Y^5}$.
\\ Following similar arguments we have
\begin{equation} \dot{Y}=\sqrt{c-\delta Y^{-4}}.\end{equation}
The dissipative term
$$ g(Y)= -\frac{\delta}{Y^3 \sqrt{cY^4-\delta}}.$$
and a parametric solution to equation (3.11) in terms of Weierstrass $\wp$ function is given as
$$
t=y_0^2\omega+\big[-\frac{y_0f^{'}(y_0)}{2\wp^{'}(\omega_0)}+\frac{f^{'}(y_0)^2\wp^{''}(\omega_0)}{16\wp^{'}(\omega_0)^3}\big]log\frac{\sigma(\omega+\hat{c}+\omega_0)}{\sigma(\omega+\hat{c}-\omega_0)}
\hskip200pt$$ $$-\frac{f^{'}(y_0)^2}{16\wp^{'}(\omega_0)^2}\big[\zeta(\omega+\hat{c}+\omega_0)+\zeta(\omega+\hat{c}-\omega_0)\big]\hskip200pt
$$
\begin{equation}
+(\omega+\hat{c})\big(\frac{y_0f^{'}(y_0)}{\wp^{'}(\omega_0)}\zeta(\omega_0)-\frac{f^{'}(y_0)^2}{16}
\big[\frac{2\wp(\omega_0)}{\wp^{'}(\omega_0)^2}+\frac{2\wp^{''}(\omega_0)\zeta(\omega_0)}
{\wp^{'}(\omega_0)^3}\big]\big)+\hat{\delta } \hskip50pt
\end{equation}
and
\begin{equation}
\displaystyle Y=y_0+\frac{f^{'}(y_0)}{4[\wp(\omega+\hat{c})-\frac{f^{''}(y_0)}{24}]}
\end{equation}
where $\hat{\delta}$ is an integrating constant and $\hat{c}$ being any fixed constant.
\\Also $f(Y)=cY^4-\delta $, $y_0$ is a root of the equation $f(Y)=0$
 and $\rho(\omega)=\rho(\omega, g_2, g_3)$ is the {\it Weierstrass} $\rho$ - {\it function} attached to the {\it Weierstrass Invariants}.
Here $g_2=3\alpha_2^2-4\alpha_1\alpha_3$ and $g_3=2\alpha_1\alpha_2\alpha_3-\alpha_2^3-\alpha_0\alpha_3^2;$
$\alpha_0=c$, $\alpha_1=cy_0 $, $\alpha_2=cy_0^2 $, $\alpha_3=cy_0^3. $
$\sigma(\omega)$ and $\eta(\omega)$ are the  Weierstrass sigma and  Weierstrass zeta functions respectively, $\wp(\omega_0)=\frac{f^{''}(y_0)}{24}$ (for a choice of $\omega_0$ ). Expression (3.12) is obtained from formula 1037.11 in \cite{BF}.

\bigskip

Generalizing the results of equation (2.1)
we consider the following example of generalized damped Milne-Pinney equation.
Let us consider the equation
\begin{equation}
\displaystyle\ddot{Y}+g(Y)\dot{Y}+\lambda Y =\frac{k_1}{Y^3}+\frac{k_2}{Y^2}+\sum_{n=0}^R\delta_n Y^{2n+1}
\end{equation}
For $R=0$ we have
\begin{equation}
\displaystyle\dot{Y}= \sqrt{2(\delta_0-\lambda)Y^2+c-4k_2Y^{-1}-2k_1Y^{-2}}
\end{equation}
The dissipative term
 $$g(Y)=\frac{(\lambda-\delta_0)Y-k_2Y^{-2}-k_1Y^{-3}}{\sqrt{2(\delta_0-\lambda)Y^2+c-4k_2Y^{-1}-2k_1Y^{-2}}}$$
Further integration of (3.15) yields a parametric solution in terms of Weierstrass $\wp$ function given as
\begin{equation}
t=y_0\omega+\frac{f^{'}(y_0)}{4\wp^{'}(\omega_0)}[\log\frac{\sigma(\omega+\hat{c}-\omega_0)}{\sigma(\omega+\hat{c}+\omega_0)}
+2(\omega+\hat{c})\zeta(\omega_0)]+\delta
\end{equation}

\begin{equation}
\displaystyle Y=y_0+\frac{f^{'}(y_0)}{4[\wp(\omega+\hat{c})-\frac{f^{''}(y_0)}{24}]}
\end{equation}
where $\delta$ is an integrating constant and $\hat{c}$ being any fixed constant.
\\Also $f(Y)=2(\delta_0-\lambda)Y^4+cY^2-4k_2Y-2k_1 $, $y_0$ is a root of the equation $f(Y)=0$
 and $\wp(\omega)=\wp(\omega, g_2, g_3)$ is the {\it Weierstrass} $\wp$ - {\it function.} Here $g_2=3\alpha_2^2-4\alpha_1\alpha_3$ and $g_3=2\alpha_1\alpha_2\alpha_3-\alpha_2^3-\alpha_0\alpha_3^2;$
$$\alpha_0=2(\delta_0-\lambda)\hskip4000pt$$
$$\alpha_1=2(\delta_0-\lambda)y_0\hskip400pt$$
$$\alpha_2=2(\delta_0-\lambda)y_0^2+\frac{c}{6}\hskip400pt$$
$$\alpha_3=2(\delta_0-\lambda)y_0^3+3\frac{cy_0}{6}-k_2\hskip400pt.$$
\\ $\wp(\omega_0)=\frac{f^{''}(y_0)}{24}$ (for a choice of $\omega_0$ )is not equal to any of the roots of $4y^3-g_2y-g_3=0$. Expression (3.16) is obtained from formula 1037.06 in \cite{BF}.
\section{Conclusion}
In this letter we have computed the solutions to a class of exactly integrable generalized damped Milne-Pinney equations.  If the
coefficients of the second-order nonlinear equations satisfy some specific conditions that follow from the Chiellini
integrabilty, then the general solution of the damped Milne-Pinney equation can be obtained in an exact parametric form.
In particular, we have obtained the solutions of the damped Ermakov-Painlev\'e II and another generalized damped Milne-Pinney equation.

\end{document}